\documentclass[aps,reprint,superscriptaddress,nofootinbib,floatfix,longbibliography]{revtex4-2}

\usepackage{amsmath}
\usepackage{bm}
\usepackage{amsfonts}
\usepackage{graphicx}
\usepackage{hyperref}
\usepackage{xcolor}

\enlargethispage{\baselineskip}

\hypersetup{
     colorlinks   = true,
     citecolor    = blue
}
\everymath{\displaystyle}

\newcommand{\TDLI}{\affiliation{Tsung-Dao Lee Institute (TDLI) \& School of Physics and Astronomy, Shanghai Jiao Tong University, \\ Shengrong Road 520, 201210 Shanghai, P.\ R.\ China}}

\begin{document}

\title{Electroweak axion in light of GRB221009A}

\author{Weikang Lin}
\email{linweikang-academia@outlook.com}
\affiliation{South-Western Institute For Astronomy Research, Yunnan University, Kunming 650500, Yunnan, P. R. China}
\TDLI

\author{Tsutomu T. Yanagida}
\email{tsutomu.tyanagida@sjtu.edu.cn }
\TDLI
\affiliation{Kavli IPMU (WPI), The University of Tokyo, Kashiwa, Chiba 277-8583, Japan}

\begin{abstract}
   Very high energy (VHE) photons may have a higher survival rate than that expected in standard-model physics, as suggested by the recently reported Gamma Ray Burst GRB221009A. While a photon-axion like particle (ALP) oscillation can boost the survival rate of the VHE photons, current works have not been based on concrete particle models, leaving the identity of the corresponding ALP unclear. Here, we show that the required ALP scenario is consistent with the electroweak axion with an anomaly free $Z_{10}$ Froggatt-Nielsen symmetry. 
\end{abstract}

\maketitle

Recently, an unexpectedly bright and long-duration Gamma Ray Burst (GRB), GRB221009A, has been detected at redshift $z=0.15$ (corresponding to a comoving distance of $\sim 600$ Mpc) \cite{GCN32632,*GCN32635,*GCN32637,*GCN32648}. LHAASO reported more than 5000 VHE photons with energies up to $\sim18$ TeV associated with this event \cite{LHASSO32677}. Later, Carpet-2 reported a $251$ TeV photon-like air shower \cite{Atel15669}.

The detection of VHE photons at such a distance is difficult to explain with conventional physical processes. In particular, due to the interaction with the extragalactic background light (EBL) via an electron-positron pair production, VHE photons have a large optical depth and their survival rate is extremely low \cite{Franceschini:2017iwq}. 

While the GRB221009A anomaly of VHE photons still needs close examinations, it is soon pointed out in \cite{Galanti:2022pbg,Baktash:2022gnf} that the survival rate can be significantly increased by a photon-ALP oscillation. The VHE photons may convert back and forth to ALPs in the presence of a magnetic field in the host galaxy, the intergalactic medium and the Milky Way. During the ALP phase, they evade the interaction with EBL, and hence the optical depth is reduced. With the benchmark ALP mass $m_A=10^{-10}$ eV and photon-ALP coupling constant $g_{A\gamma\gamma}=0.5\times10^{-11}$ GeV$^{-1}$, the survival rate for photons with $E\gtrsim10^{13}$ eV is significantly boosted \cite{Galanti:2022pbg}. This benchmark value of $g_{A\gamma\gamma}=0.5\times 10^{-11}$ is consistent with the constraints from both the star cooling effects (see \cite{Galanti:2022ijh} and references wherein) and laboratory experiments \cite{CAST:2017uph}. The benchmark ALP mass is within the region of ALP mass ($m_A>10^{-12}$ eV) that allows such a high $g_{A\gamma\gamma}$ \cite{Galanti:2022ijh}.

However, the suggested ALP mass is far smaller than that of the QCD axion given $g_{a\gamma}\simeq10^{-11}$ GeV$^{-1}$. In this short paper, we point out that the suggested parameter space is consistent with the electroweak (EW) axion whose mass is generated by the electroweak instantons \cite{Fukugita:1994xn, Nomura:2000yk}.

The EW axion is a hypothetical pseudo-Nambu-Goldstone boson field that couples to the $SU(2)$ EW gauge fields with the following interaction, 
\begin{equation}\label{eq:AWW-coupling}
    \mathcal{L}\supset\frac{g^2_2}{32\pi^2} \frac{A}{F_A} W_{\mu \nu}^i\widetilde{W}^{i\mu \nu}\,,
\end{equation}
where $W_{\mu \nu}^i$ (with $i=1,2,3$) is the weak $SU(2)$ gauge field tensor and $\widetilde{W}^{i\mu \nu}$ is its dual tensor \footnote{We consider that the Planck scale $M_{Pl} \simeq 2.4\times 10^{18}$ GeV is the cut-off scale of the theory as explained in \cite{Choi:2021aze}.}. The EW axion was originally proposed to explain why the age of the universe appeared to be shorter than that of some old stars \cite{Fukugita:1994xn}. The potential generated by the EW instanton effect is comparable to the magnitude of dark energy density observed \cite{Fukugita:1994xn,Nomura:2000yk}.\footnote{It is also shown to be able to explain the recently observed cosmic birefringence \cite{Choi:2021aze,Lin:2022niw}.} The version considered in \cite{Nomura:2000yk} is a supersymmetric model and the EW instanton effect generates the axion potential, which reads
\begin{equation}\label{eq:axio-potential}
     V_A=\frac{\Lambda_{A}^4}{2}\big(1-\cos(A/F_A)\big) \,,
\end{equation}
with
\begin{equation}\label{eq:potential-height}
\begin{split}
    \Lambda_{A}^4&\simeq2e^{-\frac{2\pi}{\alpha_2(M_{Pl})}}c\,\epsilon^{10}m_{3/2}^3M_{Pl}\\
    &\simeq c\big(\frac{\epsilon}{1/17}\big)^{10}\big(\frac{m_{3/2}}{1\,{\rm TeV}}\big)^3(1.4\times10^{-3}\,{\rm eV})^4,
\end{split}
\end{equation}
where $c$ is a dimensionless constant of $\mathcal{O}(1)$, $\epsilon \simeq 1/17$ is the suppression factor due to the Froggatt-Nielsen $U(1)_{FN}$ flavor symmetry, $m_{3/2}$ is the gravitino mass and $F_A$ is the decay constant.\footnote{This result does not change even if some $SU(2)$ charged particles exist at the intermediate energy scale owing to the SUSY miracle \cite{Nomura:2000yk}.  }
The axion potential Eq.\,\eqref{eq:axio-potential} gives us the EW axion mass $m_A$ around the potential minimum as
\begin{equation}\label{eq:axion-mass}
     m_A=\frac{\Lambda_A^2}{\sqrt{2}F_A}\simeq\big(\frac{\epsilon}{1/17}\big)^{10}\big(\frac{m_{3/2}}{1\,{\rm TeV}}\big)^{\frac{3}{2}}\big(\frac{M_{Pl}}{F_A}\big) \times6\times10^{-34}\,{\rm eV}\,,
\end{equation}
taking $c=1$.

\begin{figure}
    \centering
    \includegraphics[width=\linewidth]{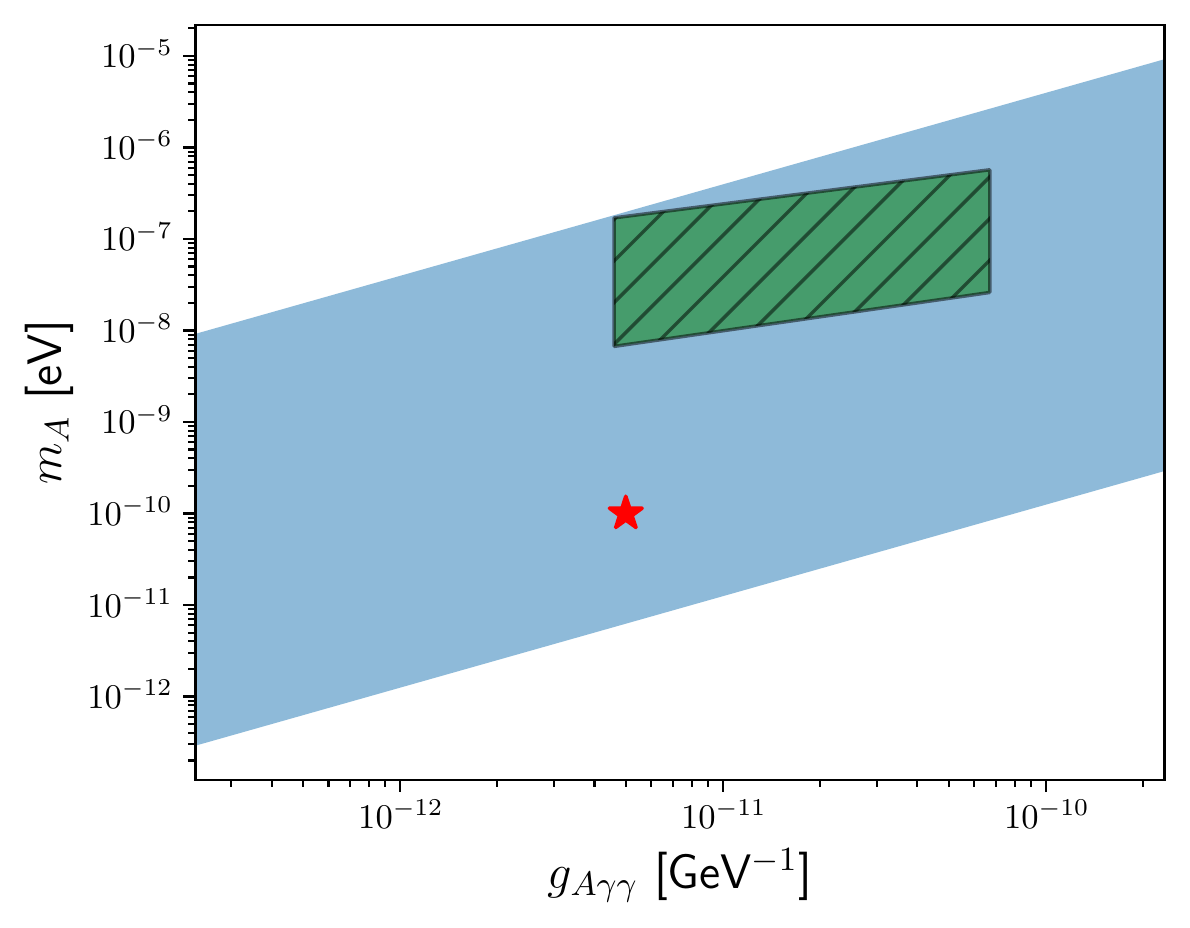}
    \caption{Parameter space in the EW axion model with an anomaly free $Z_{10}$ Froggatt-Nielsen symmetry. The shaped band corresponds to that with a range of the gravitino mass $m_{3/2}=1-1000$ TeV. The red star denotes the benchmark parameters adopted in \cite{Galanti:2022pbg}. The green and hatched} region represents the constraint given in \cite{Troitsky:2022xso}.
    \label{fig:ma_gayy}
\end{figure}

The EW axion has, in general, a coupling to the electromagnetic gauge fields as
\begin{equation}\label{eq:couplingAFF}
    \mathcal{L} \supset -c_\gamma\frac{\alpha}{4\pi}\frac{A}{F_A} F_{\mu\nu}\tilde{F}^{\mu\nu}\,,
\end{equation}
where $\alpha\simeq1/137$ is the fine structure constant, $c_\gamma$ is an anomaly coefficient, $F_{\mu\nu}$ and $\tilde{F}^{\mu\nu}$ are the Faraday tensor and its dual. The EW axion interaction to the weak gauge field in Eq. \eqref{eq:axio-potential} itself leads to the above axion-photon coupling after the electroweak phase transition. In this minimal model, $c_\gamma=1$  \cite{Choi:2021aze} which we assume below.\footnote{While $c_\gamma$ is defined via Eq.\,\eqref{eq:coupling-constant}, it can have other contributions in some models. It is in a minimal model as considered here that we have $c_\gamma=1$. } We identify the EW axion-photon coupling constant as 
\begin{equation}\label{eq:coupling-constant}
    g_{A\gamma\gamma} =\frac{c_\gamma\alpha}{\pi F_A}\,.
\end{equation}
The benchmark value $g_{A\gamma\gamma}=0.5\times 10^{-11}$ GeV$^{-1}$ corresponds to $F_A =5\times 10^8$ GeV, which gives too small an EW axion mass $m_A\simeq 10^{-22}$ eV for $m_{3/2}=10$ TeV compared to the required $m_A\simeq10^{-10}$ eV taken in \cite{Galanti:2022pbg}.  

However, the above discussion has another option of flavor symmetry. Above, We have assumed the Froggatt-Nielsen $U(1)_{FN}$ flavor symmetry. This symmetry must be a global symmetry since it has gauge anomalies. However, a discrete $Z_{10}$ subgroup of the $U(1)_{FN}$ is anomaly free and can be a gauge symmetry \cite{Choi:2019jck, Lin:2022khg}. If we assume this anomaly-free $Z_{10}$ gauge symmetry, we can remove the suppression factor $\epsilon^{10}$. In that case, the EW axion mass is given by \cite{Choi:2019jck, Lin:2022khg},
\begin{equation}\label{eq:EWaxionMass}
    m_A\simeq\big(\frac{m_{3/2}}{1\,{\rm TeV}}\big)^{\frac{3}{2}}\big(\frac{M_{Pl}}{F_A}\big) \times1.2\times10^{-21}\,{\rm eV}\,.  
\end{equation}
Then, for $g_{A\gamma\gamma}=0.5\times 10^{-11}$ GeV$^{-1}$ (again corresponding to $F_A =5\times 10^8$ GeV), we get an EW axion mass $m_A\simeq 10^{-10}$ eV for $m_{3/2}\simeq7$ TeV,  which is well consistent with the benchmark ALP mass taken in \cite{Galanti:2022pbg}. We show the predicted mass-coupling constant relation of the EW axion in Fig. \ref{fig:ma_gayy} taking the gravitino mass $m_{3/2} = 1-1000$ TeV which is consistent with experimental constraints \cite{Ibe:2011aa,Arkani-Hamed:2012fhg}. For the relevant range of $m_A$, this value of $g_{A\gamma\gamma}$ (and hence $F_A$) is allowed by the constraint $g_{A\gamma\gamma}<0.66\times 10^{-10}$ from the CERN Axion Solar Telescope (CAST) \cite{CAST:2017uph}.

We have shown the ALP scenario that boosts the survival rate of VHE photons is consistent with the EW axion. Since the required coupling constant $g_{A\gamma\gamma}$ is only about one order of magnitude lower than the constraint from CAST that searches for hypothetical ALPs emitted from the Sun \cite{CAST:2017uph}, we may be able to detect the EW axion by the next-generation solar axion experiment such as IAXO \cite{Irastorza:2011gs,Armengaud:2014gea,IAXO:2019mpb,IAXO:2020wwp}. It should be pointed out that a significant portion of parameter space of this version of EW axion, especially the parameter space required to explain the GRB221009A anomaly obtained in another analysis \cite{Troitsky:2022xso} can be probed by the future Any Light Particle Search (ALPS-II) laboratory experiment \cite{Bahre:2013ywa}.

\textbf{Note added:} After our work appeared on arXiv, we found \cite{Troitsky:2022xso} which explores the ALP parameter space that explains the increased survival rate of the VHE photos. While their result is somehow different from the benchmark parameters adopted in \cite{Galanti:2022pbg} due to different scenarios considered, it is consistent with the predicted mass-coupling constant relation of the EW axion. We show in addition the constraint from \cite{Troitsky:2022xso} to Fig. \ref{fig:ma_gayy}. Note that the lower bound of $m_A$ would be smaller if the intergalactic magnitude field is smaller \cite{Troitsky:2022xso}.

\begin{acknowledgments}
W. L. thanks Zheng Wang for the clarification of observed properties of GRB 221009A. We thank Ariel Arza for useful discussion on the future detection. T. T. Y. is supported in part by the China Grant for Talent Scientific Start-Up Project and by Natural Science Foundation of China (NSFC) under grant No. 12175134 as well as by World Premier International Research Center Initiative (WPI Initiative), MEXT, Japan.
\end{acknowledgments}


\bibliographystyle{apsrev4-1}
\bibliography{refs}

\end{document}